\documentclass[12pt,fleqn]{article}
\usepackage{amsbsy,amsmath,amssymb}
\usepackage{cite,graphicx}

\newcommand*{\be}{\begin{equation}}
\newcommand*{\ee}{\end{equation}}
\newcommand*{\ba}{\begin{array}}
\newcommand*{\ea}{\end{array}}
\newcommand*{\bea}{\begin{eqnarray}}
\newcommand*{\eea}{\end{eqnarray}}
\newcommand*{\bean}{\begin{eqnarray*}}
\newcommand*{\eean}{\end{eqnarray*}}

\newcommand*{\lp}{\left(}
\newcommand*{\rp}{\right)}
\newcommand*{\ls}{\left[}
\newcommand*{\rs}{\right]}
\newcommand*{\lc}{\left\{}
\newcommand*{\rc}{\right\}}

\newcommand*{\la}{\langle}

\newcommand*{\ra}{\rangle}

\renewcommand*{\d}{\textrm{d}}
\newcommand*{\e}{\mathrm{e}}
\newcommand*{\veps}{\varepsilon}

\newcommand*{\kB}{k_{\rm B}}

\newcommand*{\ds}{\displaystyle}

\newlength{\glength}
\newcommand*{\g}{\makebox[\glength][c]{\text{\slshape g}}}

\newcommand*{\CPL}{Chem. Phys. Lett.\ }
\newcommand*{\JACS}{J. Am. Chem. Soc.\ }
\newcommand*{\JCP}{J. Chem. Phys.\ }
\newcommand*{\JML}{J. Mol. Liquids\ }

\newcommand*{\JPC}{J. Phys. Chem.\ }

\newcommand*{\MP}{Mol. Phys.\ }

\newcommand*{\PRE}{Phys. Rev. E\ }

\begin{document}
\vspace*{2cm}
\noindent
{\large\bf
Pressure dependence of diffusion co\-ef\-fi\-ci\-ent
and orientational relaxation time for acetonitrile and methanol in water:
DRISM/\-mo\-de-co\-up\-ling study
}
\vspace*{1.5ex}

\noindent{\sc A.E.Kobryn$^1$, T.Yamaguchi$^2$ and F.Hirata$^{1,}$%
\footnote{To whom correspondence should be addressed. E-mail: hirata@ims.ac.jp}
}
\vspace*{1.5ex}

{\noindent\footnotesize
$^1$Institute for Molecular Science, Myodaiji, Okazaki, Aichi 444-8585, Japan\\
$^2$Department of Molecular Design and Engineering, Graduate School of Engineering,\\
\phantom{$^2$}Nagoya University, Chikusa, Nagoya, Aichi 464-8603, Japan}
\vspace*{3ex}

{\noindent\small We present results of theoretical description and
numerical calculation of the dynamics of molecular liquids based on the
Reference Interaction Site Model / Mode-Coupling Theory. They include the
tem\-pe\-ra\-tu\-re-pres\-su\-re(den\-si\-ty) dependence of the
translational diffusion coefficients and orientational relaxation times
for acetonitrile and methanol in water at infinite dilution. Anomalous
behavior, i.e. the increase in mobility with density, is observed for the
orientational relaxation time of methanol, while acetonitrile does not
show any deviations from the usual. This effect is in qualitative
agreement with the recent data of MD simulation and with experimental
measurements, which tells us that presented theory is a good candidate to
explain such kind of anomalies from the microscopical point of view and
with the connection to the structure of the molecules.}

\subsection*{Introduction}

Dynamics in solutions has been a central issue in the physical chemistry.
It has been so because liquid dynamics is an important factor which
determines the rate of chemical processes in solution including chemical
reactions and protein folding. A variety of experimental methods have
been devised to observe liquid dynamics, which now cover the dynamic
range from femtoseconds to days.

Theoretical studies of liquid dynamics and relaxation have been dominated
during the past century by the two continuum models which had been
established essentially in the 18th century: the electrodynamics and the
hydrodynamics \cite{stokes}. The monumental achievements made in the
field of liquid dynamics by the great scientists such as A.~Einstein
\cite{einstein}, P.~Debye \cite{debye}, L.~Onsager \cite{onsager,hubbard}
and M.~Born \cite{born-friction} are essentially based on the continuum
theories: the dielectric relaxation, the ion mobility and its
concentration dependence, and so on. Such theories represented by the
Stokes-Einstein-Debye (SED) law rely on the boundary conditions as well
as phenomenological use of macroscopic constants such as the viscosity
and dielectric constants in order to realize ``chemistry'' specific of
the problem in concern. Typical examples of the boundary conditions are
the ``slip'' and ``stick'' ones often employed in equations based on the
hydrodynamic theory.

The continuum models have played crucial roles in describing physics
taking place in solution with the spatial as well as temporal resolutions
which are low enough so that molecular details of the process can be
neglected. There have been gaps between the phenomenological model and
molecular processes. The gaps have been filled by using such adjectives
as ``effective''. For example, the conductivity of a simple ion in
aqueous solutions at infinite dilution depends on their size in entirely
opposite manner to what is predicted from the Stokes law, if one employs
the crystal radii as the ion size. In order to fill the gap between the
phenomenological model and the molecular process, people have invented an
``effective'' radius or the Stokes radius, and interpreted the phenomenon
in terms of the enhanced Stokes radii due to ``hydration''. So called
``fractional exponent'', e.g. \cite{Zwanzig85}, often used in the
interpretation of the viscosity dependence of a relaxation rate in
liquids is another example which tries to compromise the contradiction
between phenomenological models and experiments. If an experimentalist
tries to plot a relaxation rate against viscosity, he will often face
serious violation of a law predicted by a phenomenological theory. It has
been a common maneuver in such cases to use ``fractional'' dependence on
viscosity. However, the use of ``fractional'' dependence does not add any
information to describe molecular process actually taking place in
solutions, but just hides the breakdown of the phenomenological theory.
Recent experimental techniques, which are dramatically improved both in
resolution and sensitivity, have demonstrated unambiguously breakdown of
the phenomenological model. Let us refer to two contributions which
exhibit such breakdown. Nakahara \cite{nakahara} and co-workers have
shown that the rotational dynamics of a solute in a variety of solvents
does not correlate with the viscosity of the solvent but rather with
specific interactions between solute and solvent. Terazima and his
coworkers \cite{terazima} have found that the diffusion constant of
radical species is roughly a half of that of their parent molecules,
which are about the same size. We suppose no more words are necessary to
convince people what may happen if one tries to describe more complicated
dynamics in solution, e.g., conformational change of protein and chemical
reactions, in terms of the phenomenological models.

Are there, then, any hope for theories of solution dynamics to break
through the old regime established by the great scientists? We say
``yes'', if one relies on the two theories in the statistical mechanics
developed in the last century: the generalized Langevin equation (GLE)
\cite{hansen} and the RISM theory of molecular liquids
\cite{Chandler72,Chandler82,xrism}. The generalized Langevin theory
describes the time evolution of few dynamic variables in concern as a
function of the representative point in the phase space. All other
variables which are not under explicit concern are projected onto the
dynamic variables of interest with the help of a projection operator. The
projection leads to an equation which looks similar to the
phenomenological Langevin equation containing the frictional force
proportional to the rate of change of the dynamic variables, and the
random force, which are related to each other by a
fluc\-tu\-a\-ti\-on-dis\-si\-pa\-ti\-on theorem. If one chooses as the
dynamic variables the density and the conjugated flux of sites or atoms
of molecules in liquids, the theory gives an equation for the dynamic
structure factor of atoms, which describes the time evolution of the
site-site density pair correlation functions \cite{sssv}. Results from
the RISM theory, the site-site static structure factor and the direct
correlation functions, are naturally plugged into GLE in order to
evaluate not only the initial condition of the dynamics or the static
structure factor, but also the frictional force as well as the collective
frequency which concerns the frequency of the site-site density
fluctuation. A crucial development of the theory is rather conceptual,
not mathematical, in the sense that it has provided a new concept to view
dynamics of  a molecule in solution, which is quite different from the
model traditionally exploited in the field. The new model sees the
molecular motion in liquid as a correlated {\itshape translational}
motion of atoms: if two atoms in a diatomic molecule are moving in the
same direction, then the molecule as a whole is making a translational
motion, while the molecule should be rotating if its atoms  are moving in
opposite directions. The view is different from the
rotational-translational model traditionally developed based on the
angular coordinates \cite{calef,bagchi,wei}.

The new theory of liquid dynamics has been successfully applied to a
variety of relaxation process in solution including the collective
excitations in water \cite{chong-water},  ion dynamics in dipolar liquids
\cite{chong-friction}, dynamical Stokes shift \cite{sssv-tcf,nishiyama},
pressure dependence of the transport coefficients \cite{yamaguchi2} as
well as dielectric relaxation spectrum in water \cite{yamaguchi1}, and so
forth. In the present proceeding, we report our latest study on dynamics
of a molecule in solution at infinite dilution and its temperature and
density dependence.

\subsection*{Theory}
\subsubsection*{Equilibrium structure}

In this work, we use the dielectrically-consistent reference
interaction-site model (or DRISM for brevity) formalism
\cite{Perkyns92} for the system. The main equation here is the
site-site Orn\-stein-Zer\-ni\-ke equation (SSOZ) written as
\begin{equation}
\boldsymbol{\rho}\tilde{\mathbf{h}}(k)\boldsymbol{\rho}-\tilde{\boldsymbol{\chi}}(k)=
[\tilde{\boldsymbol{\omega}}(k)+\tilde{\boldsymbol{\chi}}(k)]\tilde{\mathbf{c}}(k)
[\tilde{\boldsymbol{\omega}}(k)+\tilde{\boldsymbol{\chi}}(k)]+
[\tilde{\boldsymbol{\omega}}(k)+\tilde{\boldsymbol{\chi}}(k)]\tilde{\mathbf{c}}(k)
[\boldsymbol{\rho}\tilde{\mathbf{h}}(k)\boldsymbol{\rho}-\tilde{\boldsymbol{\chi}}(k)],
\end{equation}
where $\boldsymbol{\rho}$ is the diagonal matrix of number density
of molecular species, and $\tilde{\mathbf{h}}(k)$,
$\tilde{\mathbf{c}}(k)$ and $\tilde{\boldsymbol{\omega}}(k)$ are
the total, direct and intramolecular correlation matrices,
respectively, in the reciprocal space. The matrix
$\tilde{\boldsymbol{\chi}}(k)$ is determined by the dielectric
properties of the system.

In the case of infinite dilution limit, i.e. when concentration of
one species (called a solute) tends to zero and concentration of
other (called a solvent) essentially determines the total density,
the DRISM equation can be decomposed as follows:
\begin{subequations}
\label{SSOZ}
\begin{eqnarray}
\tilde{\mathbf{h}}^{\mathrm{vv}}(k)&=&
[\tilde{\mathbf{w}}^{\mathrm{v}}(k)+\tilde{\mathbf{D}}^{\mathrm{v}}(k)\boldsymbol{\rho}^{\mathrm{v}}]
\tilde{\mathbf{c}}^{\mathrm{vv}}(k)
[\tilde{\mathbf{w}}^{\mathrm{v}}(k)+\boldsymbol{\rho}^{\mathrm{v}}\tilde{\mathbf{h}}^{\mathrm{vv}}(k)]
+\tilde{\mathbf{D}}^{\mathrm{v}}(k),\\
\tilde{\mathbf{h}}^{\mathrm{uv}}(k)&=&\tilde{\mathbf{w}}^{\mathrm{u}}(k)\tilde{\mathbf{c}}^{\mathrm{uv}}(k)
[\tilde{\mathbf{w}}^{\mathrm{v}}(k)+\boldsymbol{\rho}^{\mathrm{v}}\tilde{\mathbf{h}}^{\mathrm{vv}}(k)],\\
\tilde{\mathbf{h}}^{\mathrm{uu}}(k)&=&
\tilde{\mathbf{w}}^{\mathrm{u}}(k)\tilde{\mathbf{c}}^{\mathrm{uu}}(k)\tilde{\mathbf{w}}^{\mathrm{u}}(k)+
\tilde{\mathbf{w}}^{\mathrm{u}}(k)\tilde{\mathbf{c}}^{\mathrm{uv}}(k)\boldsymbol{\rho}^{\mathrm{v}}\tilde{\mathbf{h}}^{\mathrm{vu}}(k),
\end{eqnarray}
\end{subequations}
where
$\tilde{\mathbf{w}}(k)=\tilde{\boldsymbol{\omega}}(k)\boldsymbol{\rho}^{-1}$,
$\tilde{\mathbf{D}}(k)=\boldsymbol{\rho}^{-1}\tilde{\boldsymbol{\chi}}(k)\boldsymbol{\rho}^{-1}$,
and superscripts ``u'' and ``v'' refer to ``solute'' and
``solvent'', respectively. Equations (\ref{SSOZ}) are solved with
the hypernetted chain (HNC) type of closure specified as
\begin{eqnarray}
\g_{\alpha\gamma}(r)=\exp\left[-\frac{\phi_{\alpha\gamma}(r)}{\kB T}
+h_{\alpha\gamma}(r)-c_{\alpha\gamma}(r)\right],
\label{HNC}
\end{eqnarray}
where functions without tilde are those in the real space,
$\alpha$ and $\gamma$ are site labels, and
$\g_{\alpha\gamma}(r)\equiv[h_{\alpha\gamma}(r)+1]$ is the
site-site radial distribution function. Interaction potential
between sites $\alpha$ and $\gamma$ is denoted as
$\phi_{\alpha\gamma}(r)$, $\beta=1/\kB T$ with $\kB$ and $T$ being
the Boltzmann constant and the absolute temperature, respectively.

\subsubsection*{The site-site mode-coupling theory}

The site-site intermediate scattering function of neat solvent and its
self-parts are obtained by the generalized Langevin equation / modified
mo\-de-co\-up\-ling theory for molecular liquids as is described in the
previous study by Yamaguchi {\itshape{et al.}} \cite{yamaguchi2}.
Generalized Langevin equations for the neat solvent are given by
\begin{subequations}
\begin{eqnarray}
\ddot{\mathbf{F}}(k,t)+k^2\mathbf{J}(k)\cdot\mathbf{S}^{-1}(k)
\cdot\mathbf{F}(k,t)+\int_0^t\d\tau\;\mathbf{K}(k,t-\tau)
\cdot\dot{\mathbf{F}}(k,\tau)&=&\mathbf{0},\label{GLE}\\
\ddot{\mathbf{F}}^{\mathrm{s}}(k,t)+k^2\mathbf{J}(k)\cdot\mathbf{w}^{-1}(k)
\cdot{\mathbf{F}}^{\mathrm{s}}(k,t)+\int_0^t\d\tau\;\mathbf{K}^{\mathrm{s}}(k,t-\tau)
\cdot\dot{\mathbf{F}}^{\mathrm{s}}(k,\tau)&=&\mathbf{0}.\label{eq:GLE_self}
\end{eqnarray}
\end{subequations}
Here, the site-site intermediate scattering function and its self-part in
the time domain, denoted as $\mathbf{F}(k,t)$ and
$\mathbf{F}^{\mathrm{s}}(k,t)$, respectively, are given by
\begin{subequations}
\begin{eqnarray}
F_{\alpha\gamma}(k,t)&\equiv&\frac1N
\langle\rho^*_\alpha(k,t=0)\rho_\gamma(k,t)\rangle,\\
F^{\mathrm{s}}_{\alpha\gamma}(k,t)&\equiv&\frac1N
\langle\rho^*_\alpha(k,t=0)\rho_\gamma(k,t)\rangle^{\mathrm{s}},
\end{eqnarray}
\end{subequations}
where $N$ is the number of solvent molecules, $\rho_\alpha(k)$ is the
density field of $\alpha$-site in the reciprocal space, and
$\la\cdots\ra$ means the statistical average, the suffix ``s'' means that
correlations between the quantities of different {\itshape molecules} are
neglected. Connection with equilibrium properties is expressed in the
relation between the site-site static structure factor
$\mathbf{S}(k)\equiv\mathbf{F}(k,t=0)$ and direct and intramolecular
correlation functions, which is
\begin{eqnarray}
S_{\alpha\gamma}(k)&\equiv&\frac1N\la\rho^*_\alpha(k,t=0)\rho_\gamma(k,t=0)\ra
=\left[\tilde{\mathbf{w}}^{\mathrm{v}}(k)
+\rho\tilde{\mathbf{h}}^{\mathrm{vv}}(k)\right]_{\alpha\gamma}.
\end{eqnarray}
The site-current correlation matrix $\mathbf{J}(k)$ is defined in a
similar way as
\begin{eqnarray}
J_{\alpha\gamma}(k)\equiv\frac1N
\langle j^*_{\alpha,z}(k,t=0)j_{\gamma,z}(k,t=0)\rangle,
\end{eqnarray}
where $z$-axis is taken to be parallel to the $\mathbf{k}$ vector. The
mode-coupling expressions of the memory function matrices, denoted as
$\mathbf{K}(k,t)$ and $\mathbf{K}^{\mathrm{s}}(k,t)$, are given by
\begin{subequations}
\begin{eqnarray}
\left[\mathbf{J}^{-1}(k)\cdot\mathbf{K}_{\mathrm{MCT}}(k,t)\right]_{\alpha\gamma}
&=&\frac{\rho}{(2\pi)^3}\int\d\mathbf{q}\;\lc q_z^2\lfloor\tilde{\mathbf{c}}(q)
\cdot\mathbf{F}(q,t)\cdot\tilde{\mathbf{c}}(q)\rfloor_{\alpha\gamma}
\mathbf{F}_{\alpha\gamma}(|\mathbf{k}-\mathbf{q}|,t)\right.\nonumber\\
\lefteqn{\ds\left.{}-q_z(k-q_z)\lfloor\tilde{\mathbf{c}}(q)
\cdot\mathbf{F}(q,t)\rfloor_{\alpha\gamma}
\lfloor\mathbf{F}(|\mathbf{k}-\mathbf{q}|,t)
\cdot\tilde{\mathbf{c}}(|\mathbf{k}-\mathbf{q}|)\rfloor_{\alpha\gamma}\rc,}
\label{eq:MCT_corr}\\\relax
[\mathbf{J}^{-1}(k)\cdot\mathbf{K}_{\mathrm{MCT}}^{\mathrm{s}}(k,t)]_{\alpha\gamma}
&=&\frac{\rho}{(2\pi)^3}\int\d\mathbf{q}\;q_z^2\lfloor\tilde{\mathbf{c}}(q)
\cdot\mathbf{F}(q,t)\cdot\tilde{\mathbf{c}}(q)\rfloor_{\alpha\gamma}
\mathbf{F}^{\mathrm{s}}_{\alpha\gamma}(|\mathbf{k}-\mathbf{q}|,t).\label{eq:MCT_self}
\end{eqnarray}
\end{subequations}
According to the recipe by Yamaguchi and Hirata \cite{Yamaguchi02},
memory functions for the self-part are given by the linear combination of
the corresponding mode-coupling memory functions as
\begin{eqnarray}
\left[\mathbf{K}^{\mathrm{s}}(k,t)\cdot\mathbf{J}(k)\right]_{\alpha\gamma}
=\mathop{\sum_{m_{1,2,3}\in\{x,y,z\}}}\limits_{\mu,\nu\in i}\relax
\langle u_{zm_1}^{(i)}Z_{m_1m_2}^{\alpha\mu}Z_{m_2m_3}^{\nu\gamma}u_{zm_3}^{(i)}
\e^{i\mathbf{k}\cdot\left(\mathbf{r}_i^\alpha-\mathbf{r}_i^\mu
-\mathbf{r}_i^\gamma+\mathbf{r}_i^\nu\right)}\rangle\nonumber\\
{}\times\left[\mathbf{J}^{-1}(k)
\cdot\mathbf{K}_{\mathrm{MCT}}^{\mathrm{s}}(k,t)\right]_{\alpha\gamma},
\label{eq:Ksaxial}
\end{eqnarray}
where $Z_{m_1m_2}^{\alpha\gamma}$ and $u_{zm}^{(i)}$ stand for the
ori\-en\-ta\-ti\-on-de\-pen\-dent si\-te-si\-te velocity correlation
matrix and the unitary matrix that describes the rotation between the
molecular- and la\-bo\-ra\-to\-ry-fix\-ed coordinates of molecule $i$,
respectively \cite{Yamaguchi02}. The col\-lec\-ti\-ve-part of the memory
function (neglecting the orientational correlation between different
molecules) is given by
\begin{eqnarray}
\mathbf{K}(k,t)=\mathbf{K}_{\mathrm{MCT}}(k,t)+\mathbf{K}^{\mathrm{s}}(k,t)
-\mathbf{K}_{\mathrm{MCT}}^{\mathrm{s}}(k,t).
\end{eqnarray}

The time-evolution of the self-part of the so\-lu\-te-so\-lu\-te
si\-te-si\-te intermediate scattering function is governed by the
equations similar to Eqs. (\ref{eq:GLE_self}), (\ref{eq:MCT_self}) and
(\ref{eq:Ksaxial}) as
{\setlength{\arraycolsep}{0pt}
\begin{eqnarray}
&&\ddot{\mathbf{F}}^{\mathrm{uu,s}}(k,t)+k^2\mathbf{J}^{\mathrm{u}}(k)
\cdot\mathbf{w}^{\mathrm{u},-1}(k)
\cdot\mathbf{F}^{\mathrm{uu,s}}(k,t)+\int_0^t\d\tau\;
\mathbf{K}^{\mathrm{uu,s}}(k,t-\tau)
\cdot\dot{\mathbf{F}}^{\mathrm{uu,s}}(k,\tau)=\mathbf{0},\\\relax
&&[\mathbf{J}^{\mathrm{uu},-1}(k)\cdot\mathbf{K}_{\mathrm{MCT}}^{\mathrm{uu,s}}(k,t)]_{\alpha\gamma}
=\frac{\rho}{(2\pi)^3}\int\d\mathbf{q}\;q_z^2\lfloor\tilde{\mathbf{c}}^{\mathrm{uv}}(q)
\cdot{\mathbf{F}}^{\mathrm{vv}}(q,t)
\cdot\tilde{\mathbf{c}}^{\mathrm{vu}}(q)\rfloor_{\alpha\gamma}
\mathbf{F}^{\mathrm{uu,s}}_{\alpha\gamma}(|\mathbf{k}-\mathbf{q}|,t),\nonumber\\\\
&&\left[\mathbf{K}^{\mathrm{uu,s}}(k,t)\cdot\mathbf{J}^{\mathrm{uu}}(k)\right]_{\alpha\gamma}
=\mathop{\sum_{m_{1,2,3}\in\{x,y,z\}}}\limits_{\mu,\nu\in i}\relax
\langle u_{zm_1}^{(i)}Z_{m_1m_2}^{\alpha\mu}Z_{m_2m_3}^{\nu\gamma}
u_{zm_3}^{(i)}\e^{i\mathbf{k}\cdot\left(\mathbf{r}_i^\alpha-\mathbf{r}_i^\mu
-\mathbf{r}_i^\gamma+\mathbf{r}_i^\nu\right)}\rangle\nonumber\\
&&\qquad\qquad\qquad\qquad\qquad
{}\times\left[\mathbf{J}^{\mathrm{uu},-1}(k)
\cdot\mathbf{K}_{\mathrm{MCT}}^{\mathrm{uu,s}}(k,t)\right].
\end{eqnarray}}

\subsubsection*{Diffusion coefficient and reorientational relaxation time}

Based on the Green-Kubo formula, the translational diffusion coefficient
$D$ is obtained as \cite{Chong97,Chong01}
\begin{eqnarray}
D=\frac13\int_0^\infty\d t\;Z_{\alpha\gamma}(t),
\label{eq:KuboGreen}
\end{eqnarray}
where $\mathbf{Z}(t)$ is the site-site velocity autocorrelation function,
which is described in terms of the self-part of the intermediate
scattering function $\mathbf{F}^{\mathrm{s}}(k,t)$ as
\begin{eqnarray}
Z_{\alpha\gamma}(t)\equiv\frac1N\sum_i
\langle\mathbf{v}_\alpha(0)\cdot\mathbf{v}_\gamma(t)\rangle^{\mathrm{s}}
=-\lim_{k\to0}\;\frac3{k^2}\,\ddot{F}^{\mathrm{s}}_{\alpha\gamma}(k,t).
\end{eqnarray}
In Eq. (\ref{eq:KuboGreen}), $\alpha$ and $\gamma$ refer to any two sites
in a molecule which are bound by chemical bonds. The expression for the
single-particle reorientational time is also described in terms of the
site-site velocity autocorrelation function
\cite{yamaguchi2,Chong01,YamaguchiJML}. In this work, we restrict our
discussion to the rank-1 reorientational relaxation of the dipole moment
$\boldsymbol{\mu}$ given by
\begin{equation}
\boldsymbol{\mu}_i=\sum_\alpha z_\alpha\mathbf{r}_i^\alpha,
\label{eq:mu_def}
\end{equation}
where $z_\alpha$ is the charge of the site $\alpha$. The first-rank
reorientational correlation function $C_{\boldsymbol{\mu}}(t)$ is defined
as
\begin{equation}
C_{\boldsymbol{\mu}}(t)\equiv\frac{\sum_i\langle\boldsymbol{\mu}_i(0)
\boldsymbol{\mu}_i(t)\rangle}{\sum_i\langle|\boldsymbol{\mu}_i^2|\rangle}.
\label{eq:Cmu_def}
\end{equation}
Substituting Eq. (\ref{eq:mu_def}) into Eq. (\ref{eq:Cmu_def}),
$C_{\boldsymbol{\mu}}(t)$ is related to the site-site velocity
autocorrelation function $\mathbf{Z}(t)$ as
%
\begin{eqnarray}
C_{\boldsymbol{\mu}}(t)=\frac{\sum_i\sum_{\alpha\gamma}z_\alpha z_\gamma
\langle\mathbf{r}_i^\alpha(t)\mathbf{r}_i^\gamma(0)\rangle}
{\sum_i\langle|\boldsymbol{\mu}_i^2|\rangle},\qquad
\ddot{C}_{\boldsymbol{\mu}}(t)=-\frac{N\sum_{\alpha\gamma}z_{\alpha}z_{\gamma}
Z^{\alpha\gamma}(t)}{\sum_i\langle|\boldsymbol{\mu}_i^2|\rangle}.
\end{eqnarray}
%
The reorientational correlation time of the 1st rank, $\tau$, is then
defined as the time-integration of $C_{\boldsymbol{\mu}}(t)$.

\subsection*{Description of the models}

We performed explicit calculations for two popular systems, i.e.
acetonitrile (CH$_3$CN) in water and methanol (CH$_3$OH) in water, both
in the case of infinite dilution. As for the structure and the
intermolecular potential of water we employed a model of the extended
simple point charge (SPC/E) \cite{Berendsen87}. We also put the
Lennard-Jones (LJ) core on the hydrogen atoms in order to avoid the
undesired divergence of the solution of the RISM integral equation. The
LJ parameters of the hydrogen atom, the depth of the well and the
diameter, are chosen to be $0.046$ kcal/mol and $0.7$ \AA, respectively.

In acetonitrile and methanol the methyl group is considered to be a
single interaction site (Me) located on the methyl carbon atom. So that
both chemical compounds consist of three sites interacting through the
pair potential \cite{Edwards84,Jorgensen86}
\begin{equation}
\phi(r_i,r_j)\equiv\phi(r_{ij})=\sum_{\alpha,\beta}^3\lc
4\epsilon_{\alpha\beta}\ls\lp\frac{\sigma_{\alpha\beta}}{r_{i\alpha,j\beta}}\rp^{12}
-\lp\frac{\sigma_{\alpha\beta}}{r_{i\alpha,j\beta}}\rp^{6}\rs
+\frac{z_{\alpha}z_{\beta}}{r_{i\alpha,j\beta}}\rc,
\label{interaction-potential}
\end{equation}
i.e., LJ plus Coulomb. Here $\alpha$ and $\beta$ label sites on molecules
$i$ and $j$;
$r_{i\alpha,j\beta}=|\mathbf{r}_{i\alpha}-\mathbf{r}_{j\beta}|$;
parameters $\epsilon_{\alpha\beta}$ and $\sigma_{\alpha\beta}$ are LJ
well-depths and LJ diameters defined as
$\epsilon_{\alpha\beta}=\sqrt{\epsilon_\alpha\epsilon_\beta}$ and
$\sigma_{\alpha\beta}=(\sigma_\alpha+\sigma_\beta)/2$, respectively, with
$\sigma_\alpha$ being the LJ diameter of a single site. Point charges for
acetonitrile were chosen to reproduce electrostatic potential obtained in
an {\itshape ab initio} calculations \cite{Edwards84}. Numerical values
of parameters of the site-site interaction potential
(\ref{interaction-potential}) and masses of sites are specified in
Table~\ref{parameters1}.
\newcommand*{\po}{\phantom{1}}
\newcommand*{\pmi}{\phantom{-}}
\setlength{\arraycolsep}{1pt}
\begin{table}[ht]
\caption{\footnotesize Parameters of the site-site interaction potential
(\ref{interaction-potential}): mass and charge are in atomic units,
$\sigma_{\mathrm{LJ}}$ in \AA, and $\epsilon_{\mathrm{LJ}}$ in
$10^{-14}{\textrm{ erg}}/{\textrm{molec}}$.}
\label{parameters1}
\vspace*{2ex}
\footnotesize
\begin{centering}
\begin{tabular}{clllllllllll}
&\multicolumn{3}{c}{water$^{\mathrm{a}}$}
&&\multicolumn{3}{c}{acetonitrile$^{\mathrm{b}}$}
&&\multicolumn{3}{c}{methanol$^{\mathrm{c}}$}\\
&~~~O&~H$_1$&~H$_2$&&~~Me&~~C&~~~N&&~~Me&~~~O&~~H\\\hline
$m$&16.0&1.008&1.008&&15.024&12.0&14.0&&15.024&16.0&1.008\\
$z$&\,-0.8476&0.4238&0.4238&&\po0.269&\po0.129&\,-0.398&&\po0.265&\,-0.7&0.435\\
$\sigma_{\mathrm{LJ}}$&\po3.16&0.7&0.7&&\po3.6&\po3.4&\po3.3&&\po3.74&\po3.03&1.0\\
$\epsilon_{\mathrm{LJ}}$&
\po1.084&0.3196&0.3196&&\po2.64&\po0.6878&\po0.6878&&\po1.4525&\po1.1943&0.3196\\
\hline
&\multicolumn{3}{c}{$^{\mathrm{a}}$Ref. \cite{Berendsen87}}&&
\multicolumn{3}{c}{$^{\mathrm{b}}$Ref. \cite{Edwards84}}&&
\multicolumn{3}{c}{$^{\mathrm{c}}$Ref. \cite{Jorgensen86}}\\
\end{tabular}\\
\end{centering}
\end{table}

\noindent Information about bond length can be deduced from Cartesian
coordinates $(x,y,z)$ of sites indicated in Table~\ref{parameters2}
(principal axes and the origin can be taken arbitrarily).
\begin{table}[ht]
\caption{\footnotesize Cartesian coordinates of sites.}
\label{parameters2}
\vspace*{2ex}
\footnotesize
\begin{centering}
\begin{tabular}{cclcclcc}
\multicolumn{2}{c}{water}&&\multicolumn{2}{c}{acetonitrile}&&\multicolumn{2}{c}{methanol}\\
\cline{1-2}\cline{4-5}\cline{7-8}
site&($x$,$y$,$z$), \AA&&site&($x$,$y$,$z$), \AA&&site&($x$,$y$,$z$), \AA\\
O\,&(0,\pmi0\phantom{.8165},-0.0646)&&Me&(0,\po0,\pmi1.46)&&Me&(-1.4246,\po0\phantom{.8961},\po0)\\
H$_1$&(0,\pmi0.8165,\pmi0.5127)&&C&(0,\po0,\pmi0\phantom{.17})&&O&(\pmi0\phantom{.4246},\po0\phantom{.8961},\po0)\\
H$_2$&(0,-0.8165,\pmi0.5127)&&N&(0,\po0,-1.17)&&H&(\pmi0.3004,\po0.8961,\po0)\\
\cline{1-2}\cline{4-5}\cline{7-8}
\end{tabular}\\
\end{centering}
\end{table}
In calculations, the temperature of the system is varied from $258.15$ to
$373.15$ K, and the number-density from $0.03008$ to $0.04011$
molecules/\AA${}^3$, where the number-density of water at the ambient
conditions is $0.03334$ molecules/\AA${}^3$. Connection of these
parameters with thermodynamic pressure is shown in
Table~\ref{density-pressure} (except for the metastable regions where we
do not have reliable data).
\vspace*{-1ex}
\begin{table}[ht]
\caption{\footnotesize Density-pressure correspondence for water
according to \protect\cite{Wagner}. Pressure is given in MPa.}
\label{density-pressure} \vspace*{2ex} \footnotesize
\begin{centering}
\begin{tabular}{cccccc}
$\rho$, g/cm$^3$&$n$, \AA$^{-3}$&273.15 K&298.15 K&373.15 K\\\hline
0.9&0.03008&---&---&---\\
1.0&0.03334&0.4085&6.6914&100.64\\
1.1&0.03676&257.20&296.20&451.51\\
1.2&0.04011&689.03&760.76&993.38\\\hline
\end{tabular}\\
\end{centering}
\end{table}
Tem\-pe\-ra\-tu\-re/den\-si\-ty dependent dielectric per\-mit\-ti\-vi\-ty
$\varepsilon$ for water used in numerical calculations has been evaluated
as a physical solution of an empirical nonlinear equation presented in
\cite{LandoltBornstein80}:
\begin{equation}
\veps-\frac12\lp1+\frac1{\veps}\rp=\frac1v
\lp17+\frac{9.32\cdot10^4\lp1+\frac{153}{v\cdot{T}^{1/4}}\rp}{\lp1-\frac3v\rp^2T}\rp,
\end{equation}
where $v$ is a molar volume in units of cm$^3$/mol, and $T$ is
thermodynamic temperature~in~K. This equation was also used in such
tem\-pe\-ra\-tu\-re/den\-si\-ty points where no experimental data exist.

\subsection*{Numerical methods}

From the generalized Langevin equation / modified mode-coupling theory
and the DRISM /HNC integral equation theory, the diffusion coefficients
and the reorientational relaxation times of solute molecules in solution
can be obtained based solely on the information about molecular shapes,
inertia parameters, intermolecular interaction potentials, temperature
and density. First, we calculated the site-site static structure factor
by the DRISM/HNC equation using the intermolecular interaction, molecular
shape, temperature and density. In order to improve the convergence of
the RISM calculation, we used the method of the modified direct inversion
in an iterative space proposed by Kovalenko {\itshape et al.}
\cite{Kovalenko99}. From the static site-site structure factor, we
calculated the site-site intermediate scattering function using the
site-site generalized Langevin equation / modified mode-coupling theory.
The generalized Langevin equation is time-integrated numerically. The
time-development of the correlation functions in the $k\to0$ limit is
separately treated by the analytical limiting procedure of the
theoretical expressions. In the numerical procedure, the reciprocal space
is linearly discretized as $k=(n+\frac12)\Delta k$, where $n$ is the
integer from 0 to $N_k-1$. Values of $\Delta k$ and $N_k$ are $0.061$
\AA${}^{-1}$ and 512, respectively. The diffusion coefficient $D$ was
calculated from the asymptotic slope of the time dependence of the mean
square displacement and the orientational relaxation time $\tau$ was
analyzed using the rotational autocorrelation functions.

\subsection*{Results and discussion}

\begin{figure}[htb]
\begin{centering}
\includegraphics*[bb=74 67 293 764,angle=-90,width=\textwidth]{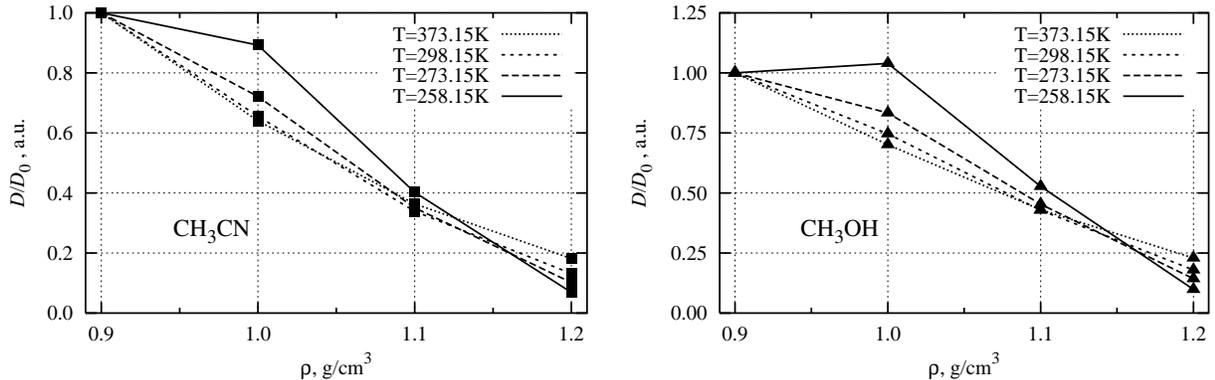}\\
\end{centering}
\caption{\footnotesize Density dependence of diffusion coefficients for
acetonitrile ({\tiny$\blacksquare$}) and methanol ($\blacktriangle$) in water at
different temperatures.}
\label{fig-d}
\end{figure}

\begin{figure}[ht]
\begin{centering}
\includegraphics*[bb=71 68 519 766,angle=-90,width=\textwidth]{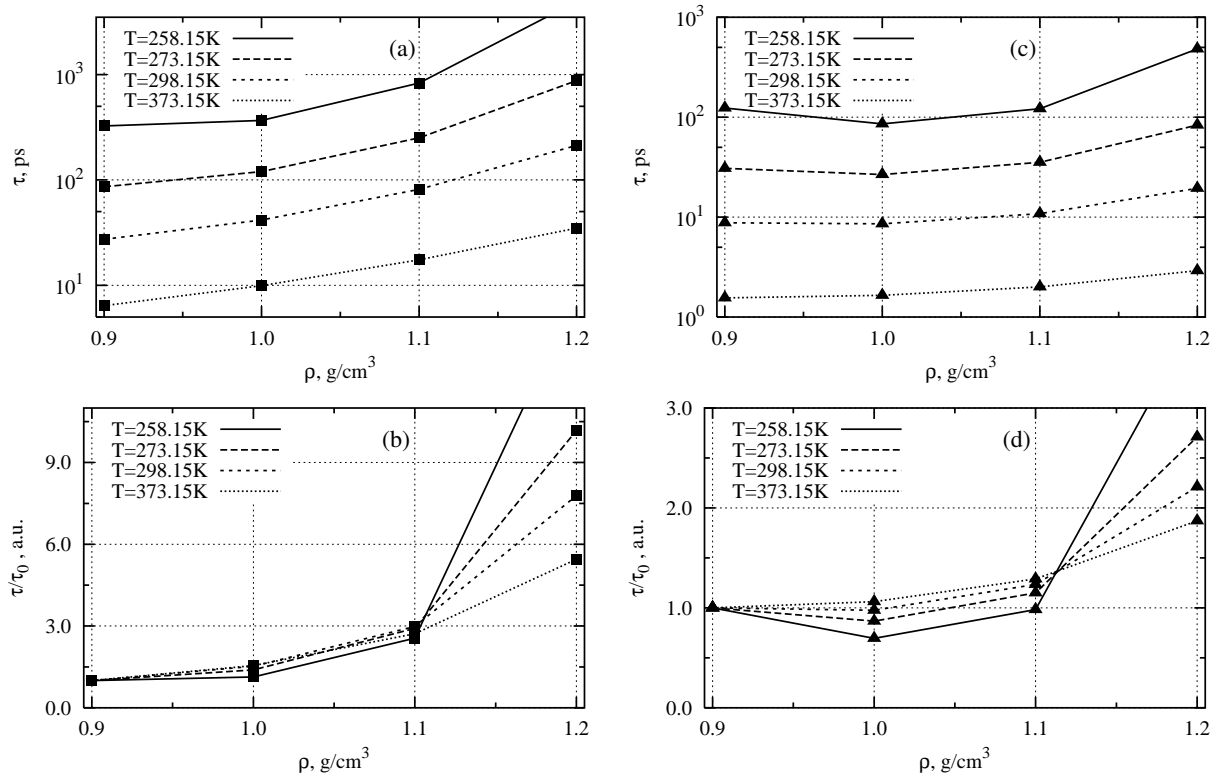}\\
\end{centering}
\caption{\footnotesize Density dependence of the orientational relaxation
times for acetonitrile ({\tiny$\blacksquare$}) and methanol ($\blacktriangle$)
in water at different temperatures.}
\label{fig-rt}
\end{figure}

Figure \ref{fig-d} shows the density dependence of normalized diffusion
coefficients $D/D_0$, with $D_0$ being the diffusion coefficient at the
lowest density, for both acetonitrile (left part) and methanol (right
part) at four temperatures. Corresponding results for the orientational
relaxation time $\tau$ are shown in figure \ref{fig-rt}. It has been
found that diffusion coefficients of both acetonitrile and methanol at
these temperatures decrease monotonically with increase of density except
for methanol at the lowest temperature, where one observes a typical
anomalous behavior: first diffusion coefficient increases, then decreases
again. At the same time, behavior of ori\-en\-ta\-ti\-o\-nal relaxation
time is different for acetonitrile and methanol at all studied
temperatures, except at the highest. One can easily observe that $\tau$
for acetonitrile ex\-hi\-bits the monotonic increase with density and
monotonic decrease of its absolute value with temperature [figure
\ref{fig-rt}(a)], while $\tau$ for methanol at the same temperatures
first decreases with density and then increases [figure \ref{fig-rt}(c)].
Its absolute value is also getting smaller with temperature as in the
case of acetonitrile. In such a way, the density dependent orientational
relaxation time for dissolved in water methanol has an absolute minimum
somewhere in between 0.9 g/cm$^3$ and 1.1 g/cm$^3$ for water and the
depth of this minimum is seen more clearly at lower temperatures. These
tendencies are observed sharply in figures \ref{fig-rt}(b) and
\ref{fig-rt}(d) for acetonitrile and methanol, respectively, where we
plot normalized values of $\tau$ as $\tau/\tau_0$ with $\tau_0$ being the
orientational relaxation time at the lowest density.
\vspace*{-2ex}
Obtained results are in qualitative agreement with the outco\-me of the
MD simulation by Chowdhuri and Chandra \cite{Chowdhuri03} performed for
similar systems at 258 K and 298 K, and with experimental measurements by
Wakai and Nakahara \cite{nakahara} who observed likewise behavior for
several different systems, including acetonitrile, at room temperatures.

The difference in behavior of acetonitrile and methanol can be made clear
from the point of view of their molecular structure. To begin with, let
us remind that the anomalous density dependence of the molecular mobility
in water has been explained by Yamaguchi \textit{et.al.}
\cite{yamaguchi2} based on the facts of an almost spherical repulsive
core for water and the strong short-range intermolecular Coulomb
interaction called ``hydrogen bonding''. Following the idea and owing to
the models used in present manuscript, one can conclude that since
acetonitrile cannot make the hydrogen bonding and methanol does, this
fact may be a good candidate to serve as a reason or cause of
justification of the anomalous density behavior of methanol (or possibly
any protic solution) and usual one of acetonitrile (or possibly any
aprotic solution).

\subsection*{Summary}

In present work we have calculated the temperature/density dependence of
the translational diffusion coefficients and rank-1 orientational
relaxation times for acetonitrile and me\-tha\-nol in water at infinite
dilution using the site-site generalized Langevin equation / modified
mo\-de-coup\-ling theory and the DRISM/HNC theory. Calculations show
anomalous density dependence of the orientational relaxation time for
methanol in water which is consistent with the results of experimental
observation and MD simulation. On the other hand, similar computation for
acetonitrile does not exhibit deviation from the usual behavior which is
also in agreement with experiment and MD simulation. Such a difference is
explained based on the molecular structure of solvents, in particular
their ability (for methanol) and inability (for acetonitrile) to create
hydrogen bonds. More studies will be required for better understanding of
the molecular dynamics in such systems.

\def\refname{{\normalsize\bf REFERENCES}}

\end{document}